\documentclass[preprint,aps,prd,epsf,showpacs,amsmath,amssymb,graphics]{revtex4-1}
\usepackage{amsmath,amssymb,amsthm}
\usepackage{color}
\usepackage{epsfig}
\usepackage{multirow}
\usepackage{graphicx}
\usepackage{subfigure}
\usepackage{rotating}
\usepackage{amsmath}

\usepackage{changes}

\begin{document}

\title{Strong gravitational lensing of rotating regular black holes in non-minimally coupled Einstein-Yang-Mills theory}
\author{Ruanjing Zhang$^{1}$\footnote{Email: zhangruanjing@126.com},
Jiliang Jing$^{2,3}$\footnote{Email: jljing@hunnu.edu.cn},
Zhipeng Peng$^{1}$,
Qihong Huang$^{4}$}
\affiliation{$^1$Institute of Theoretical Physics, School of Science, Henan University of Technology, Zhengzhou, Henan 450001, P. R. China\\
$^2$Department of Physics, Key Laboratory of Low Dimensional Quantum Structures and Quantum Control of Ministry of Education, Synergetic Innovation Center for Quantum Effects and Applications, Hunan Normal University, Changsha, Hunan 410081, P. R. China\\
$^3$Center for Gravitation and Cosmology, College of Physical Science and Technology, Yangzhou University, Yangzhou 225009, P. R. China\\
$^4$School of Physics and Electronic Science, Zunyi Normal University, Zunyi, Guizhou 563006, P. R. China}

\begin{abstract}	
The strong gravitational lensing of a regular and rotating magnetic black hole in non-minimally coupled Einstein-Yang-Mills theory is studied. We find that, with the increase of any characteristic parameters of this black hole, such as the rotating parameter $a$, magnetic charge $q$ and EYM parameter $\lambda$, the angular image position $\theta_{\infty}$ and relative magnification $r_m$  decrease while deflection angle $\alpha(\theta)$ and image separation $s$ increase. The results will degenerate to that of the  Kerr case, R–N case with magnetic charge and Schwarzschild case when we take some specific values for the black hole parameters. The results also show that, due to the small influence of magnetic charge and Einstein-Yang-Mills parameters, it is difficult for current astronomical instruments to tell this black hole apart from a General Relativity one.
\end{abstract}

\maketitle

\section{Introduction}
In recent years, with the announcement of gravitational waves \cite{Abbott2016,Abbott2016(2),Abbott2017,Abbott2017(2)} and image of black hole (BH) \cite{EHT,EHT1,EHT2,EHT3,EHT4,EHT5,EHT6}, the general relativity (GR) achieves great success and related research has attracted people's much attention \cite{Jing,Jing1,Jing2023,Jing2021,Fang,Chen,Pan}.  The gravitational lensing is also a prophecy of GR \cite{Einstein} which caused by deflection of light rays in a gravitational field and first observed by astronomer Walash et al. in 1979 \cite{Walash}. At this stage, the mainstream gravitational lensing observation projects are Kilo-Degree Survey \cite{kids}, Gaia Gravitational Lenses group \cite{gaia}, STRong-lensing Insights into Dark Energy Survey \cite{strides}, and so on. This shows that gravitational lensing is still a very active research area, because gravitational lensing can infer cosmological constants, search for planets outside our solar system, help astronomers study how our universe originated and star formed in galaxies \cite{Hanson,Courbin}, etc. This has opened a whole new avenue for researches, so we can understand the universe better. A gravitational lensing effect that produces multiple images, arcs, or even Einstein rings is known as strong gravitational lensing effect.
With the improvement of observation technology, the strong gravitational lensing has become an attractive observational effect for astronomy and fundamental physics, and thus it has been widely discussed in various theories of gravity \cite{Vir,Bozza3,Bozza,Bozza2,schen3,zhang,zhang1}.

 In recent years, non-minimal theories have numerous applications to cosmology and astrophysics, they can couple a gravitational field to other fields and media by using cross terms containing the curvature tensor. These theories have been elaborated in detail for scalar and electromagnetic fields and can be divided into five classes \cite{Azam}, i.e., the coupling of scalar fields with the spacetime curvature \cite{Geenner}, non-minimal Einstein-Maxwell models \cite{Hehl}, Einstein-Yang-Mills models \cite{Hoissen}, Einstein-Yang-Mills-Higgs models \cite{Balakin}, and non-minimal Einstein-Maxwell-Axion models \cite{Balakin2}. In these five classes of non-minimally coupled theories, various aspects of star and BH physics were discussed. In this paper, we focus on the non-minimally coupled Einstein-Yang-Mills (EYM) theory where the curvature couples with the non-Abelian gauge fields \cite{Balakin3,Balakin4}. The regular spherically symmetric solution of non-minimal EYM theory with magnetic charge of Wu-Yang gauge field was presented by Balakin \cite{Balakin3}, and the weak and strong deflection gravitational lensings of it were studied in Ref.\cite{xie}. This BH with a cosmological constant was also presented by Balakin \cite{Balakin4}, and its thermodynamics properties were studied in Ref.\cite{Jawad,Jawad2}.
Jusufi extended the spherically symmetric solution to a rotating one and studied the BH shadow, the quasinormal modes for massless scalar and electromagnetic fields, and the quasiperiodic oscillations \cite{Jusufi}. The shadow and weak gravitational lensing of a rotating regular BH in a non-minimally coupled EYM theory in the presence of plasma were studied in Ref.\cite{Kala}.

Our aim here is to study the strong gravitational lensing of a regular and rotating magnetic BH in non-minimally coupled EYM theory and the structure of the present  paper is laid out as follows: in Sec.II, we introduce the BH briefly and probe the effects of the characteristic parameters on the event horizon radius, photon sphere radius and the deflection angle. In Sec.III, by supposing that the gravitational field of the supermassive BH at the centre of our Galaxy can be described by this solution, we obtain the numerical results for the main observables in the strong deflection limit. Our conclusions and discussions are presented in the last section.

\section{Deflection angle of the  rotating regular black holes in non-minimally coupled Einstein-Yang-Mills theory}

Let us now briefly review the rotating regular BHs in non-minimally coupled EYM theory.
In the first place, the action of the theory in four-dimensional spacetimes is given by \cite{Balakin3}
\begin{equation}\label{s1}
S=\frac{1}{8\pi}\int d^{4}x\sqrt{-g} \left[ R+\frac{1}{2}F^{(a)}_{\mu \nu }F^{\mu \nu(a)}
+\frac{1}{2}\mathcal{R}^{\alpha \beta \mu \nu} F^{(a)}_{\alpha \beta }F^{(a)}_{\mu \nu} \right],
\end{equation}
where $g=det(g_{\mu \nu})$  and $R$ represent the determinant of metric tensor and the Ricci scalar, respectively.
 Additionally, as a general rule, the Greek indices run from $0$ to $3$, and the Latin indices run from $1$ to $3$. From this action, a regular, static and spherically symmetric BH can be found as \cite{Balakin3}
\begin{equation}\label{s2}
ds^2=-f(r)dt^2+f(r)^{-1}dr^2+r^2d\theta^2+r^2\sin ^2\theta d\phi ^2,
\end{equation}
with
\begin{equation}
f(r)=1+\left ( \frac{r^4}{r^4+2\lambda }  \right ) \left (-\frac{2M}{r}+\frac{q^2}{r^2}\right ),
\end{equation}
in which $M$ represents the BH mass, $q$ represents the magnetic charge of the Wu-Yang gauge field. It is very important to note that $\lambda$ called as EYM parameter is equal to $\xi q^2$ with $\xi $ being the non-minimally coupled parameter between the Yang–Mills field and the gravitational field. The above metric reduces  to the form of the Reissner–Nordström (R–N) BH but with the magnetic charge instead of the electric charge when the non-minimal coupling vanishes, i.e. $\xi =0$, and the metric reduces to the Schwarzschild case for $\lambda=0$ ($\xi =0$) and $q=0$.

By using Newman–Janis algorithm, the regular, static and spherically symmetric BH can be extended to a rotating one \cite{Jusufi}

\begin{equation}
\begin{aligned}
ds^2=&-\left (1-\frac{2Y(r) r}{\Sigma}\right ) dt^2+\frac{\Sigma }{\Delta } dr^2+\Sigma d\theta ^2
\\&+\frac{\left [ (r^2+a^2)^2-a^2\Delta\sin^2\theta\right ]\sin^2\theta }{\Sigma } d\phi^2
-2a\sin^2\theta \frac{2Y(r) r}{\Sigma} dtd\phi,
\end{aligned}
\end{equation}
with
\begin{equation}
\begin{aligned}
&\Sigma =r^2+a^2\cos^2\theta,\\
&Y(r)=\frac{r(1-f(r))}{2},\\
&\Delta=r^2+a^2-\frac{r^6}{(r^4+2\lambda)} \left ( \frac{2M}{r}-\frac{q^2}{r^2} \right ) .
\end{aligned}
\end{equation}
This metric describes a regular and rotating (identified by $a$)  magnetic BH solution  with a Yang-Mills electromagnetic source in the non-minimal EYM theory.  It can reduce to Kerr BH for $\lambda=0$ and $q=0$. In this paper, we choose to work with $M = 1$ for simplicity.

\begin{figure}[htbp]
  \centering
  \includegraphics[scale=1]{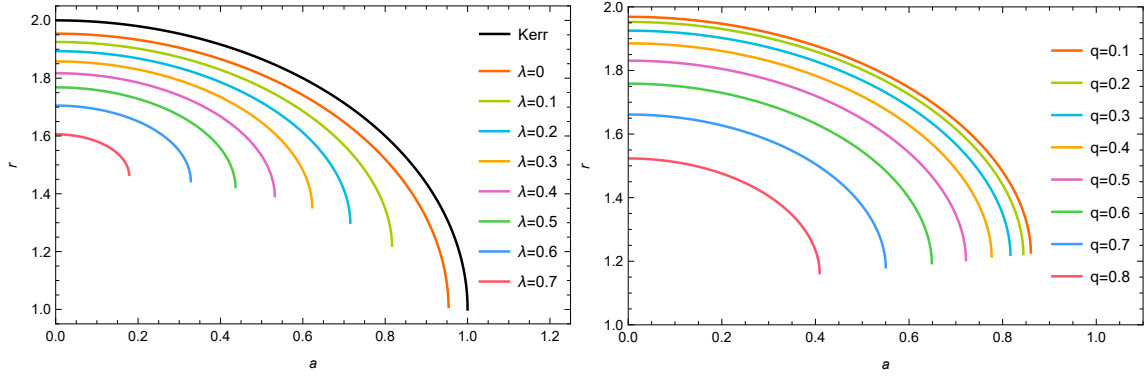}
  \caption{Variation of the event horizon radius of the regular and rotating magnetic BH with the rotating parameter $a$. The left graph is evaluated at $q=0.3$ for different $\lambda$ and right graph is evaluated at $\lambda=0.1$ for different $q$. The black line stands for the Kerr case and the intersection of black line with $a=0$ stands for the Schwarzschild case.}
  \label{rha}
  \end{figure}

\begin{figure}[htbp]
  \centering
  \includegraphics[scale=1]{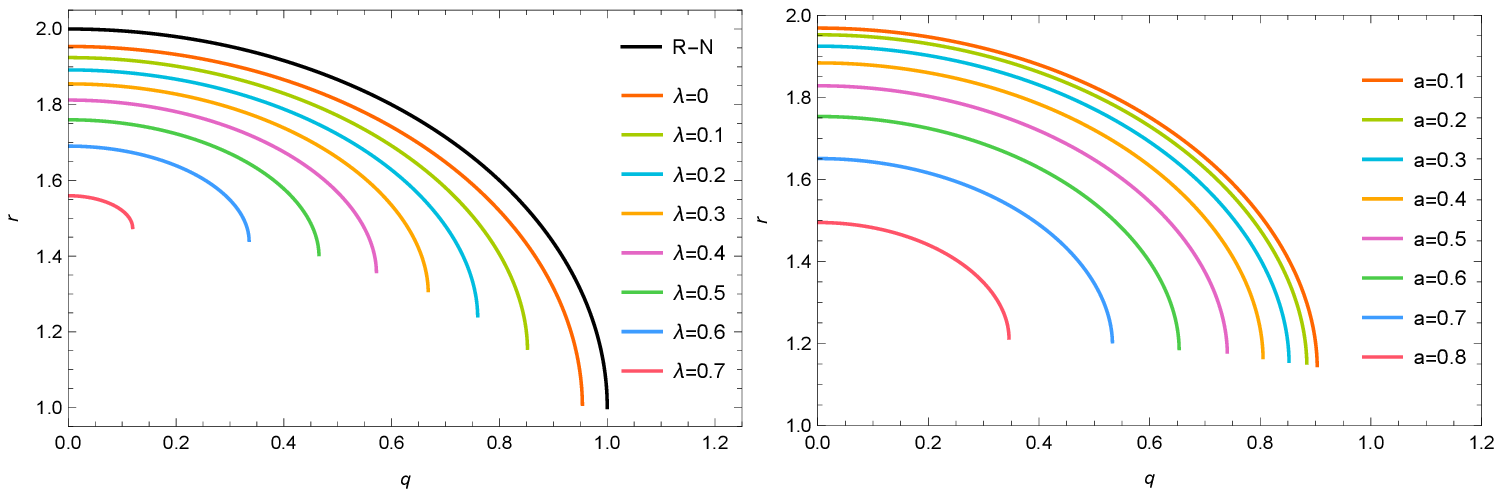}
  \caption{Variation of the event horizon radius of the regular and rotating magnetic BH with the magnetic charge $q$. The left graph is evaluated at $a=0.3$ for different $\lambda$ and right graph is evaluated at $\lambda=0.1$ for different $a$. The black line stands for the R–N case and the intersection of black line with $q=0$ stands for the Schwarzschild case.}
  \label{rhq}
  \end{figure}

\begin{figure}[htbp]
  \centering
   \includegraphics[scale=1]{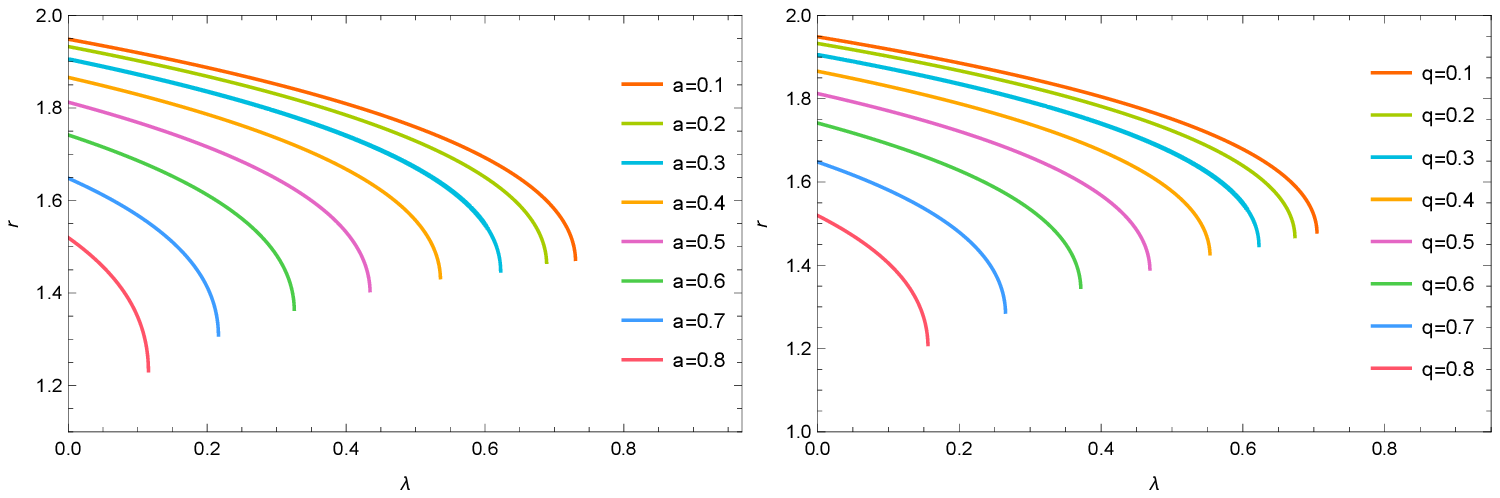}
  \caption{Variation of the event horizon radius of the regular and rotating magnetic BH with the EYM parameter $\lambda$. The left graph is evaluated at $q=0.3$ for different $a$ and right graph is evaluated at $a=0.3$ for different $q$. }
  \label{rhl}
  \end{figure}

 The position of the BH horizon can be determined by calculating $\Delta=0$ numerically. Then, the variation of event horizon radius ($r_H$) with rotating parameter $a$, magnetic charge $q$ and EYM parameter $\lambda$ is shown in Figs.(\ref{rha})–(\ref{rhl}). From these figures, we can find that the event horizon radius of BH decreases with the increase of characteristic parameters of BH. The black line in Fig.(\ref{rha}) stands for the Kerr case, the black line in Fig.(\ref{rhq}) stands for the R–N case with magnetic charge. And the values corresponding to the zero of the horizontal axis in these two black lines stand for the Schwarzschild case. These three data are consistent with previous calculations.

\begin{figure}[htbp]
  \centering
   \includegraphics[scale=1]{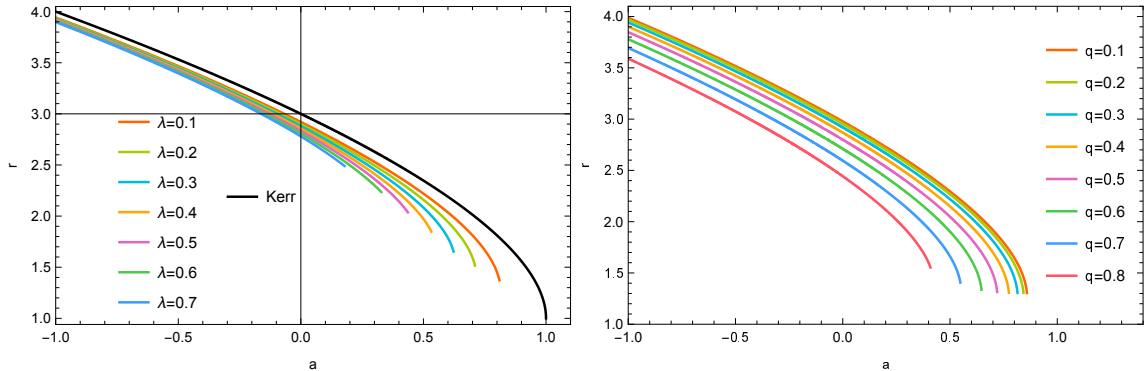}
  \caption{Variation of the photon sphere radius of the regular and rotating magnetic BH with the rotating parameter $a$. $a<0$ denotes the BH rotates in the converse direction as the photon, and $a>0$ denotes it rotates in the same direction as the photon. The left graph is evaluated at $q=0.3$ for different $\lambda$ and right graph is evaluated at $\lambda=0.1$ for different $q$. The black line stands for the Kerr case and the intersection of black line with $a=0$ stands for the Schwarzschild case.}
  \label{rpsa}
  \end{figure}

\begin{figure}[htbp]
  \centering
   \includegraphics[scale=1]{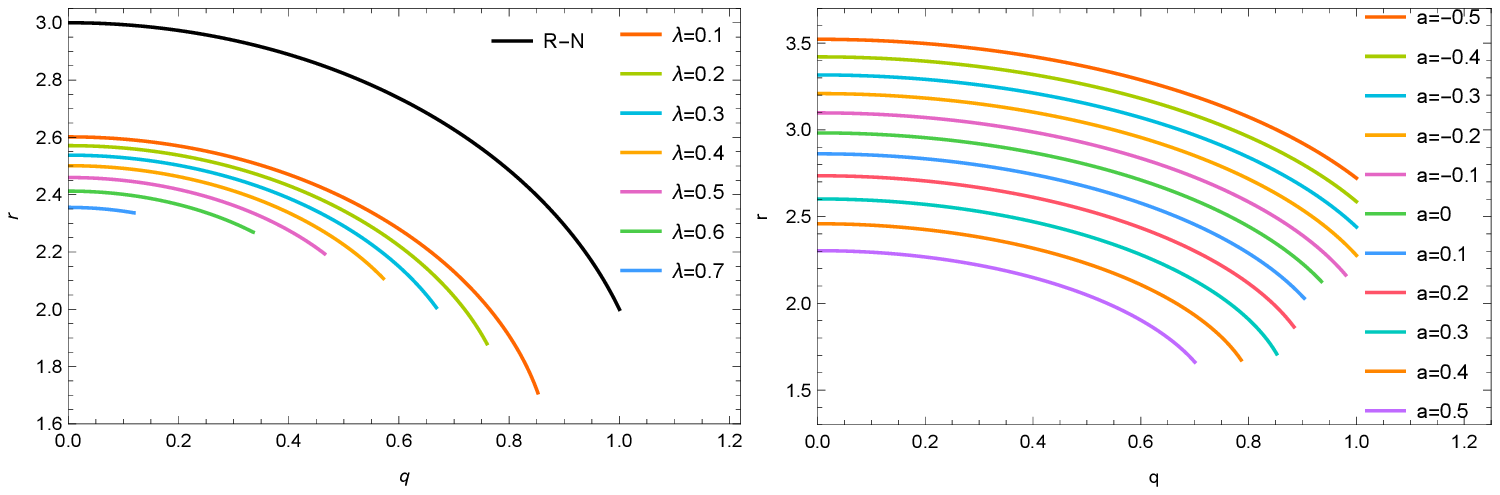}
   \caption{Variation of the photon sphere radius of the regular and rotating magnetic BH with the magnetic charge $q$. The left graph is evaluated at $a=0.3$ for different $\lambda$ and right graph is evaluated at $\lambda=0.1$ for different $a$. The black line stands for the R–N case and the intersection of black line with $q=0$ stands for the Schwarzschild case.}
  \label{rpsq}
  \end{figure}

\begin{figure}[htbp]
  \centering
   \includegraphics[scale=1]{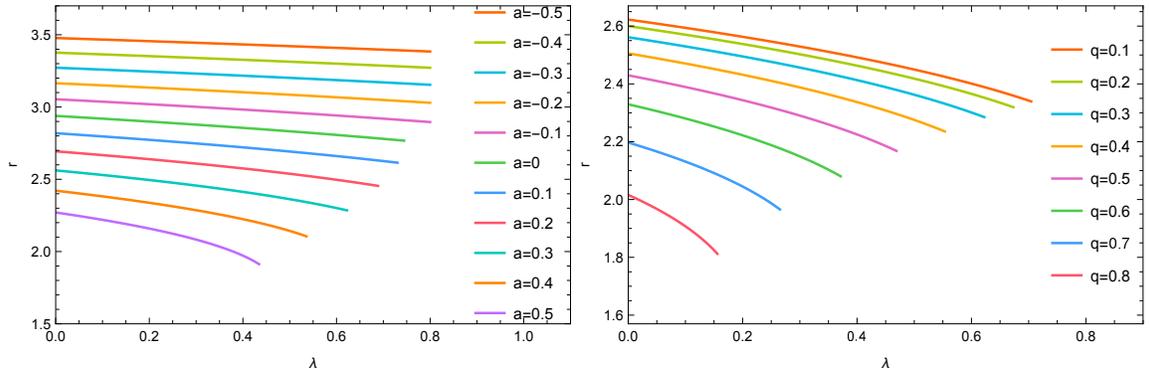}
  \caption{Variation of the photon sphere radius of the regular and rotating magnetic BH with the EYM parameter $\lambda$. The left graph is evaluated at $q=0.3$ for different $a$ and right graph is evaluated at $a=0.3$ for different $q$.}
  \label{rpsl}
  \end{figure}

In the following, the characteristics of strong gravitational lensing by the regular and rotating magnetic BH were examined. It is well-known that, in a stationary axisymmetric spacetime, if the initial position and tangent vector of a geodesic line are in the equatorial plane ($\theta=\pi/2$), then the entire geodesic line will be in that plane. Additionally, through coordinate transformation, each geodesic line can be transformed into a motion on the equatorial plane. Therefore, without loss of generality, we just consider that the whole trajectory of the photon is limited on the equatorial plane and both observer and source lie in the same plane. To simplify the calculation, the metric can be rewritten as
\begin{equation}
ds^{2}=-F(r)dt^{2}+T(r)dr^{2}+H(r)d\phi^{2}-2G(r)dtd\phi,
\end{equation}
with
\begin{equation}
\begin{aligned}
\label{abc}
&F(r)=\frac{r^4-2r^3+q^2r^2+2\lambda}{r^4+2\lambda},\\
&T(r)=\frac{r^2(r^4+2\lambda)}{r^6-2r^5+(q^2+a^2)r^4+2r^2\lambda+2a^2\lambda}, \\
&H(r)=\frac{r^6+a^2r^4+2a^2r^3-(a^2q^2-2\lambda)r^2+2a^2\lambda}{r^4+2\lambda} \\
&G(r)=-\frac{a(q^2-2r)r^2}{r^4+2\lambda}.
\end{aligned}
\end{equation}
The null geodesics for this metric are
\begin{equation}
\begin{aligned}
\frac{dt}{d\tau}&=\frac{H(r)-JG(r)}{G(r)^2+F(r)H(r)},\\
\frac{d\phi}{d\tau}&=\frac{G(r)+JF(r)}{G(r)^2+F(r)H(r)},\\
\left(\frac{dr}{d\tau}\right)^2&=\frac{H(r)-2JG(r)-J^2F(r)}{T(r)H(r)[G(r)^2+F(r)H(r)]}
\end{aligned}
\end{equation}
where $\tau$ is an affine parameter along the geodesics, and $J$, defined as angular momentum of the photon, is equal to the impact parameter $u$ in a stationary, axially-symmetric spacetime. By using the above metric, the expression of $J$ is
\begin{equation}
\begin{aligned}
\label{ju}
J&=u=\frac{-G(r_0)+\sqrt{G(r_0)^2+F(r_0)H(r_0)}}{F(r_0)}\\
&=\frac{a(q^2-2r_0)r_0^2+\sqrt{r_0^{10}-2r_0^9+(a^2+q^2)r_0^8+4\lambda r_0^6-4\lambda r_0^5+2\lambda(2a^2+q^2)r_0^4+4\lambda^2r_0^2+4a^2\lambda^2}
}{r_0^4-2r_0^3+q^2r_0^2+2\lambda},
\end{aligned}
\end{equation}
where $r_0$ is the minimum radial distance, defined by the distance of light's inflection point to the center of the BH, and the deflection angle of the light becomes unboundedly large as $r_0$ tends to the photon sphere radius $r_{ps}$ which can be obtained by solving
\begin{equation}
\label{rps1}
F(r)H'(r)-F'(r)H(r)+2J[F'(r)G(r)-F(r)G'(r)]=0.
\end{equation}
By replacing the parameter in Eq.(\ref{rps1}) with Eq.(\ref{abc}), we can get the equation of photon sphere for this regular and rotating magnetic BH. Since this equation is very complex, we can only get its solution numerically. We then show the variation of photon sphere radius $r_{ps}$ with rotating parameter $a$, magnetic charge $q$ and EYM parameter $\lambda$ in Figs.(\ref{rpsa})–(\ref{rpsl}). These figures are similar to the change of event horizon radius,  which decreases with the increase of characteristic parameters of BH. The black lines in Fig.(\ref{rpsa}) and Fig.(\ref{rpsq}) have the same meaning as that in Fig.(\ref{rha}) and Fig.(\ref{rhq}). However, there are also differences, the rotating parameter $a$ has no effect on event horizon radius, but it has an effect on the photon sphere radius $r_{ps}$, i.e., $r_{ps}$ for $a>0$, which denotes the BH rotates in the same direction of the photon, always less than $r_{ps}$ for $a<0$, which denotes the BH rotates in the opposite direction as the photon. In other words, retrograde light rays must keep farther from the center.

Typically, it is assumed that the distance between the photon source and the lens, as well as the distance between the observer and the lens, is much greater than the minimum radial distance.
Therefore, in a stationary, axially-symmetric spacetime, the deflection angle of a photon originating from infinity follows a specific rule, as
\begin{equation}
\alpha (r_0) =I(r_0)-\pi,
\end{equation}
with
\begin{equation}
\label{i}
I(r_0)=2\int_{r_0}^{\infty } \frac{\sqrt{T(r)|F(r_0)|}[G(r)+JF(r)]dr }
{\sqrt{G^2(r)+F(r)H(r)}\sqrt{F(r_0)H(r)-F(r)H(r_0)+2J[F(r)G(r_0)-F(r_0)G(r)]}  } .
\end{equation}
In a flat space, light travels along a straight line with a deflection angle of $0$, and the above integral is equal to $\pi$. In a curved spacetime, the deflection angle of light increases as it moves closer to the BH. When the minimum radial distance $r_{0}$ decreases to a certain point, the deflection angle reaches $2\pi$, meaning the light completes a full loop around the lens before reaching the observer. If $r_{0}$ continues to decrease, the photon will circle the BH multiple times before escaping. When $r_{0}$ equals the photon sphere radius $r_{ps}$, the deflection angle becomes infinitely large and the photon is captured by the BH. To study the behavior of the deflection angle near the photon sphere, we use Bozza's integral evaluation method \cite{Bozza}, which starts by defining
\begin{equation}
z=1-\frac{r_0}{r}.
\end{equation}
Then, we can rewrite the integral part of the deflection angle as
\begin{equation}
\label{ii}
I(r_0)=\int_{0}^{1} R(z,r_0)f(z,r_0)dz,
\end{equation}
where
\begin{equation}
R(z,r_0)=\frac{2r^2}{r_0\sqrt{H(z)}} \frac{\sqrt{T(z)|F(r_0)|}[G(z)+JF(z)]}{\sqrt{G^2(z)+F(z)H(z)}},
\end{equation}
\begin{equation}
f(z,r_0)=\frac{\sqrt{H(z)}}{\sqrt{F(r_0)H(z)-F(z)H(r_0) +2J[F(z)G(r_0)-F(r_0)G(z)] } } .
\end{equation}
Moreover, for all values of $z$ and $r_0$, the function $R(z,r_0)$ is regular while the function $f(z,r_0)$ diverges when $z\rightarrow 0$, i.e., the photon approaches the photon sphere.
For the sake of simplicity, we expand the divergence term of the integrand to the second order in $z$
\begin{equation}
f_0(z,r_0)=\frac{1}{\sqrt{x(r_0)z+y(r_0)z^2}} ,
\end{equation}
with
\begin{equation}
x(r_0)=\frac{r_0}{H(r_0)} \left \{ F(r_0)H'(r_0)-F'(r_0)H(r_0)
+2J[F'(r_0)G(r_0)-F(r_0)G'(r_0)]\right \} ,
\end{equation}
\begin{equation}
\begin{aligned}
y&(r_0) = \frac{r_0}{2H^2(r_0)}\Big[r_0 H(r_0)\big(F(r_0)H''(r_0)-F''(r_0)H(r_0) +2J[F''(r_0)G(r_0)-F(r_0)G''(r_0)]\big) \\
&+2\big(H(r_0)-r_0H'(r_0)\big)\big(F(r_0)H'(r_0)-F'(r_0)H(r_0) +2J[F'(r_0)G(r_0)-F(r_0)G'(r_0)]\big)\Big].
 \end{aligned}
\end{equation}
Now, by reviewing the equation of photon sphere (Eq.(\ref{rps1})), we can find that the coefficients $x(r_0)$ vanish and the leading term of the divergence in $f_0(z,r_0)$ is $z^{-1}$ when $r_0 = r_{ps}$. Therefore, the integral (\ref{ii}) will diverge logarithmically, and in order to solve the equation, we decompose it into two parts
\begin{equation}
\begin{aligned}
I_D(r_0)&=\int_{0}^{1} R(0,r_{ps})f_0(z,r_0)dz,\\
I_R(r_0)&=\int_{0}^{1}\left [ R(z,r_{0})f(z,r_0)-R(0,r_{ps})f_0(z,r_0) \right ] dz.
\end{aligned}
\end{equation}
Obviously, $I_D(r_0)$ contains the divergence term of the original integral, while Eq.(\ref{ii}) without the divergence is $I_R(r_0)$, and we have to solve $I_D(r_0)$ and $I_R(r_0)$ separately. Finally, the deflection angle in the strong field region can be expanded in the form
\begin{equation}
\label{angle}
\alpha (\theta )=- \bar{a} log\left ( \frac{\theta D_{OL} }{u_{ps}} -1 \right ) +\bar{b}
+\mathcal{O} \left ( u-u_{ps} \right )
\end{equation}
with
\begin{equation}
\begin{aligned}
\bar{a} &=\frac{R(0,r_{ps})}{\sqrt{y(r_{ps})} },\\
\bar{b} &=-\pi +I_R(r_{ps})+\bar{a}log\Big(\frac{2y(r_{ps})H(r_{ps})}
{u_{ps}F(r_{ps})[G(r_{ps})+JF(r_{ps})]}\Big),
\end{aligned}
\end{equation}
where $\bar{a}$ and $\bar{b}$ are called as strong field limit coefficients. The $D_{OL}$ represents the distance between the observer and the gravitational lens; $u_{ps}$ is the minimum impact parameter, which is obtained by taking the value of impact parameter $u$ (Eq.(\ref{ju})) at $r_{ps}$; $\theta$ represents the angular image separation between image and optical axis, then in small angle approximation, $u=\theta D_{OL}$. We can get more properties of strong gravitational lensing of the regular and rotating magnetic BH by studying the above deflection angle.

\begin{figure}[htbp]
 \centering
 \subfigure[ These three graphs are evaluated at $q=0.3$ for different $\lambda$]{
  \includegraphics[scale=1]{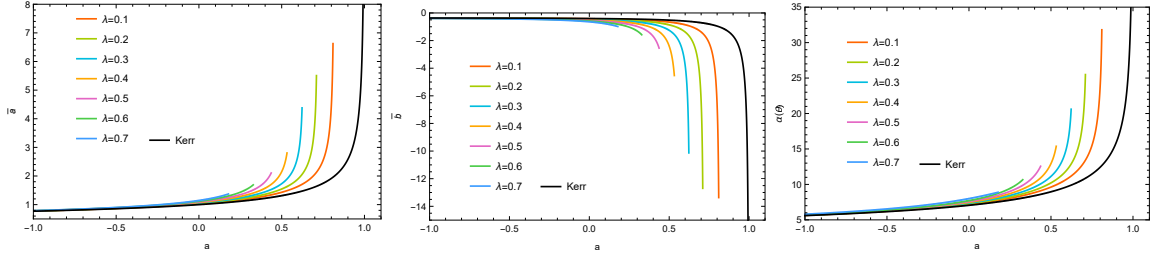}}
 \subfigure[ These three graphs are evaluated at $\lambda=0.1$ for different $q$]{ \includegraphics[scale=1]{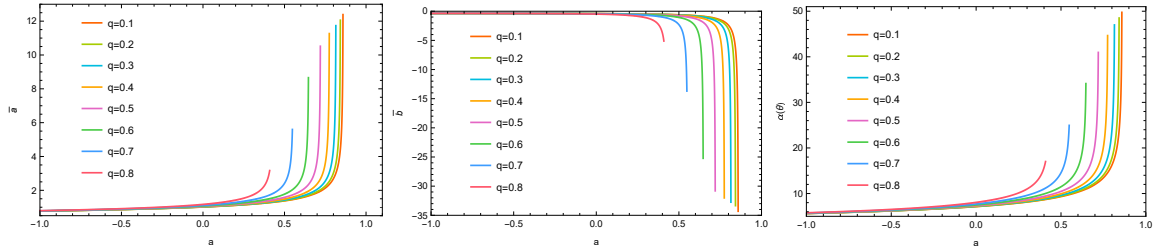}}
 \caption{Variation of the strong field limit coefficient $\bar{a}$, $\bar{b}$ and deflection angle $\alpha (\theta )$ of the regular and rotating magnetic BH with the rotating parameter $a$. The black line stands for the Kerr case and the intersection of black line with $a=0$ stands for the Schwarzschild case.}
 \label{abal}
 \end{figure}

\begin{figure}[htbp]
 \centering
 \subfigure[These three graphs are evaluated at $a=0.3$ for different $\lambda$]{
  \includegraphics[scale=1]{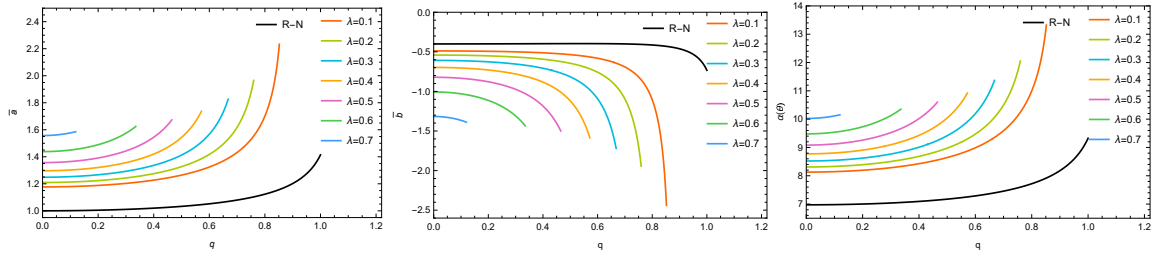}}
 \subfigure[These three graphs are evaluated at $\lambda=0.1$ for different $a$]{
  \includegraphics[scale=1]{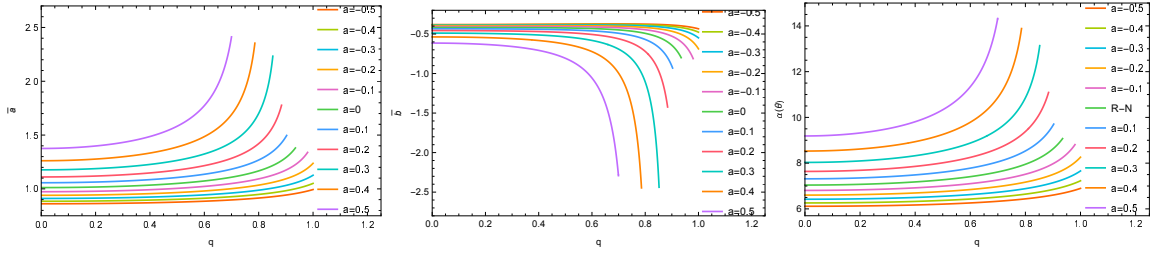}}
 \caption{Variation of the strong field limit coefficient $\bar{a}$, $\bar{b}$ and deflection angle $\alpha (\theta )$ of the regular and rotating magnetic BH with the magnetic charge $q$. The black line stands for the R–N case and the intersection of black line with $q=0$ stands for the Schwarzschild case.}
 \label{abql}
 \end{figure}

\begin{figure}[htbp]
 \centering
 \subfigure[These three graphs are evaluated at $q=0.3$ for different $a$.]{
 \includegraphics[scale=1]{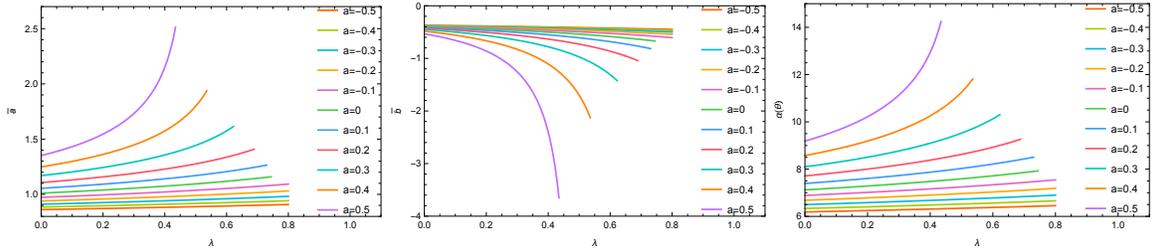}}
 \subfigure[These three graphs are evaluated at $a=0.3$ for different $q$.]{
  \includegraphics[scale=1]{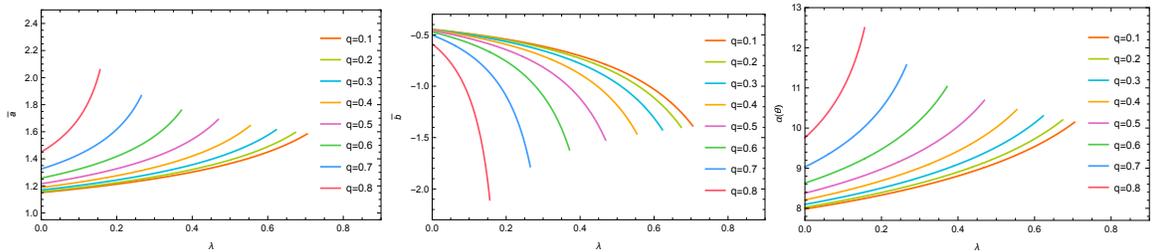}}
 \caption{Variation of the strong field limit coefficient $\bar{a}$, $\bar{b}$ and deflection angle $\alpha (\theta )$ of the regular and rotating magnetic BH with the EYM parameter $\lambda$.}
 \label{ablq}
 \end{figure}

The strong field limit coefficients $\bar{a}$ and $\bar{b}$, as well as the deflection angle $\alpha (\theta )$ evaluated at $u = u_{ps} + 0.003$ for the regular and rotating magnetic BH, are shown in Figs.(\ref{abal})–(\ref{ablq}). These three figures show that $\bar{a}$ and $\alpha (\theta )$ increase while $\bar{b}$ decreases with the increase of characteristic parameters of BH. After comparing $\bar{a}$ and $\alpha (\theta )$ in each figure, we find that their changing trends are almost exactly the same, which means that $\alpha (\theta )$ is determined by the logarithmic term in the strong field limit. Consistent with above analysis, the black lines in Fig.(\ref{abal}) and Fig.(\ref{abql}) stand for the Kerr case and the R–N case with magnetic charge, respectively, and the values corresponding to the zero of the horizontal axis in these two black lines stand for the Schwarzschild case. Now, let's turn attention to the Kerr line, where $\bar{a}$, $\bar{b}$ and $\alpha (\theta )$ all diverge at $a=1$, and it warns that the strong field limit deflection angle (Eq.(\ref{angle})) no longer represents a reliable description in the regime of high $a$. Thus, we take $a<1$ for the regular and rotating magnetic BH.
By comparing the figures of $r_h$, $r_{ps}$ and $\alpha (\theta )$, it can be seen that as the characteristic parameters of BH increase, the deflection angle increases while the event horizon radius and photon sphere radius decrease.
Especially for rotating parameter $a$, the $r_{ps}$ for anterograde light rays ($a>0$) is always smaller than the retrograde ones ($a<0$), but the change in deflection angle $\alpha (\theta )$ is opposite to $r_{ps}$. Our prior understanding that the deflection angle increases as light approaches the BH is consistent with this.

\section{Observables of strong gravitational lensing of rotating regular black holes in non-minimally coupled Einstein-Yang-Mills theory}

Now let us study the effect of rotating parameter $a$, magnetic charge $q$ and EYM parameter $\lambda$ on the observables of strong gravitational lensing. In order to describe the geometric relations among the observer, source and image, we use the lens equation as
\begin{equation}
\beta =\theta -\frac{D_{LS}}{D_{OS}} \triangle \alpha_n,
\end{equation}
where $\beta$ and $D_{LS}$ are the angular separation and distance between the lens and source, respectively, while $D_{OS}$ represents the distance between the observer and source.
Since the source, lens and observer are highly aligned, they can be contacted by $D_{OS}=D_{LS}+D_{OL}$. And $\triangle \alpha_n= \alpha(\theta) - 2n\pi$, where $n$ stands for the circle number of light. This is due to the fact that in strong gravitational lensing, light rays rotate around a high-density body multiple times before exiting, resulting in a deflection angle greater than $2\pi$. Thus, a high-density body acting as a lens can produce an endless series of relativity images on either side of the optic axis.
The first relativistic image formed by light circling the BH once is usually the outermost and most luminous, and with additional circle number, the magnification decreases exponentially. For simplicity, we assume that the first image $\theta_1$ is a single image and combine all subsequent images into one group at $\theta_{\infty}$. Finally, observables in the strong deflection lensing, i.e., angular image position $\theta_{\infty}$,  angular image separation $s$ and relative magnifications $r_m$ between $\theta_1$ and $\theta_{\infty}$ can be formulated as
\begin{equation}
\begin{aligned}
\label{sr}
\theta_{\infty}&=\frac{u_{ps}}{D_{OL}},\\
s&=\theta_{\infty}exp\left (\frac{\bar{b}-2\pi}{\bar{a} }\right ),\\
r_m&=2.5\log_{10}{\left [exp\left ( \frac{2\pi}{\bar{a}} \right ) \right]} .
\end{aligned}
\end{equation}
From Eq.(\ref{sr}), we can find that the minimum impact parameter $u_{ps}$ and the strong field limit coefficients $\bar{a}$ and $\bar{b}$ can be obtained easily by measuring the observables of relativistic images. Subsequently, by contrasting their values with the predictions of theoretical models, we can ascertain the characteristics of the BH contained in the lensing.

\begin{figure}[htbp]
 \centering
 \subfigure[ These three graphs are evaluated at $q=0.3$ for different $\lambda$]{
 \includegraphics[scale=1]{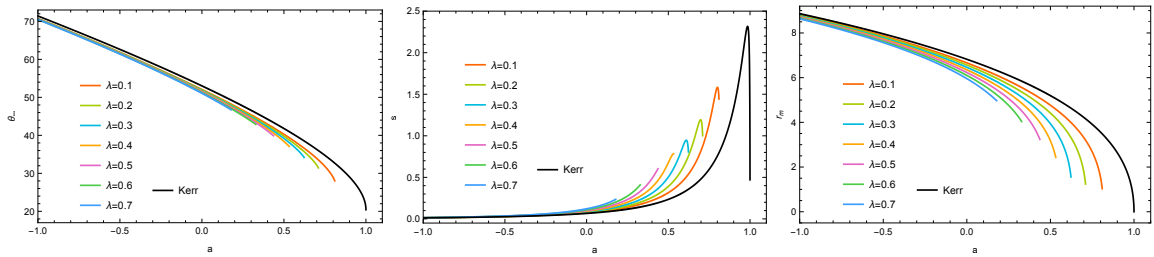}}
 \subfigure[ These three graphs are evaluated at $\lambda=0.1$ for different $q$]{ \includegraphics[scale=1]{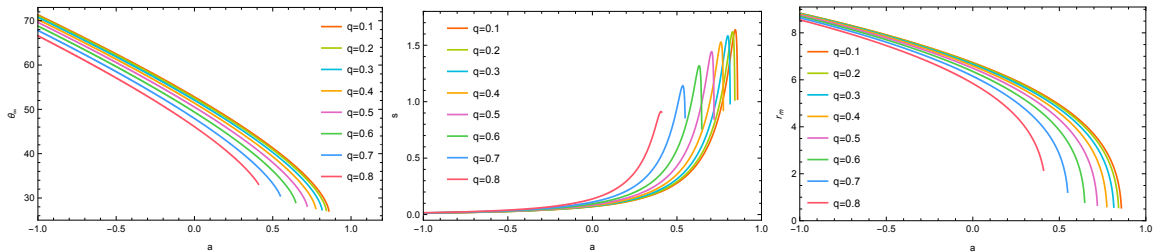}}
 \caption{Variation of the angular image position $\theta_\infty$, angular image separation $s$ and relative magnifications $r_m$ of the regular and rotating magnetic BH with the rotating parameter $a$. The black line stands for the Kerr case and the intersection of black line with $a=0$ stands for the Schwarzschild case.}
 \label{sral}
 \end{figure}

\begin{figure}[htbp]
 \centering
 \subfigure[These three graphs are evaluated at $a=0.3$ for different $\lambda$]{
  \includegraphics[scale=1]{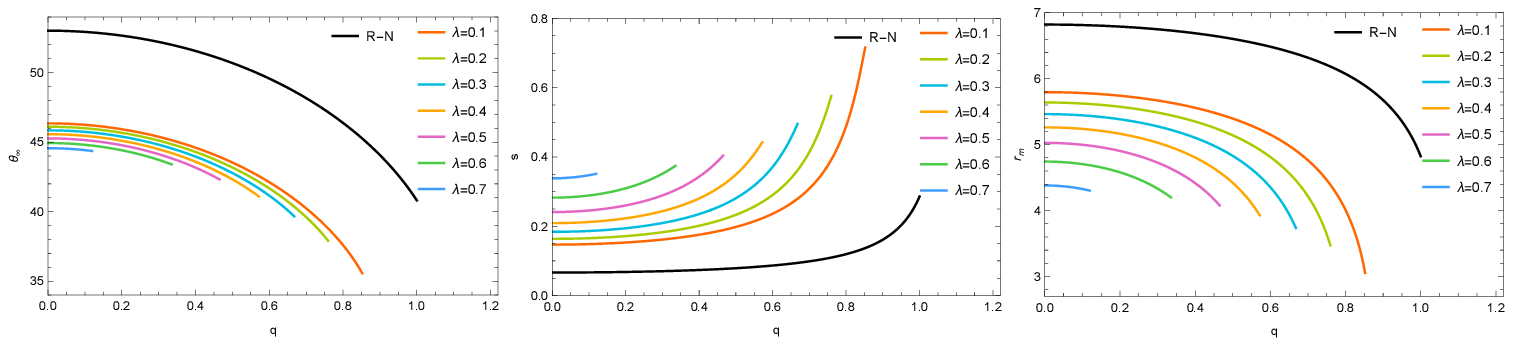}}
 \subfigure[These three graphs are evaluated at $\lambda=0.1$ for different $a$]{
 \includegraphics[scale=1]{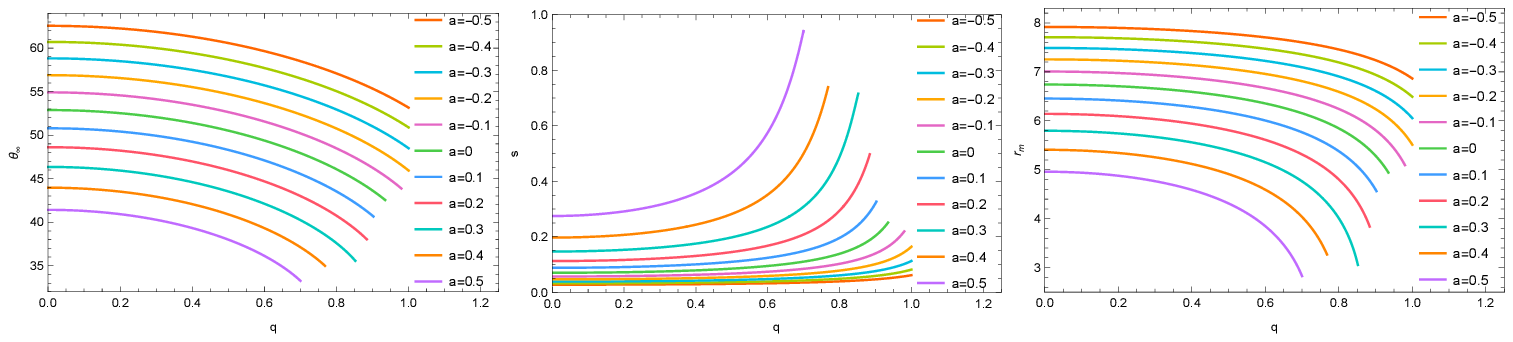}}
 \caption{Variation of the angular image position $\theta_\infty$, angular image separation $s$ and relative magnifications $r_m$ of the regular and rotating magnetic BH with the magnetic charge $q$. The black line stands for the R–N case and the intersection of black line with $q=0$ stands for the Schwarzschild case.}
 \label{srql}
 \end{figure}

\begin{figure}[htbp]
 \centering
 \subfigure[These three graphs are evaluated at $q=0.3$ for different $a$.]{
 \includegraphics[scale=1]{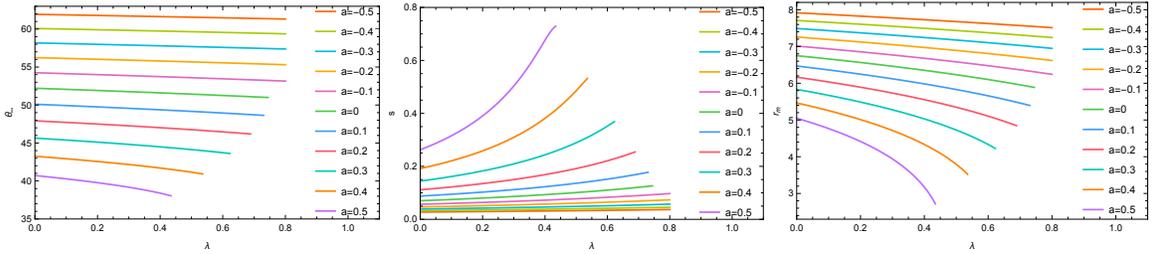}}
 \subfigure[These three graphs are evaluated at $a=0.3$ for different $q$.]{
 \includegraphics[scale=1]{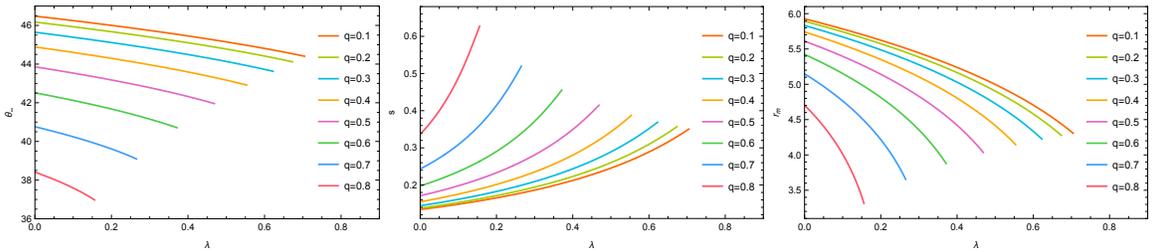}}
 \caption{Variation of the angular image position $\theta_\infty$, angular image separation $s$ and relative magnifications $r_m$ of the regular and rotating magnetic BH with the EYM parameter $\lambda$.}
 \label{srlq}
 \end{figure}

In the case of Galactic, the observables in strong deflection lensing are easily measurable. Therefore, we assume that the lens is the supermassive BH at the center of Galactic, characterized by a regular and rotating magnetic BH with the EYM parameter $\lambda$. The BH's mass is estimated to be $M=4.4\times 10^6 M_\odot$ and it is located $8.5 kpc~(D_{OL})$ from the  Earth \cite{Genzel}. Then we show the numerical values of the observables in Figs.(\ref{sral})-(\ref{srlq}). From these figures, we can find that, when any two of the three parameters (rotating parameter $a$, magnetic charge $q$ and EYM parameter $\lambda$) are fixed and one parameter increases, angular image position $\theta_\infty$ and relative magnification $r_m$ will always decrease monotonically, while angular image separation $s$ will increase. Moreover, from Fig.(\ref{sral}), it is interesting to find that, no matter what values $q$ and $\lambda$ are taken, the angular image separation $s$ can increase to a peak value for a certain $a$. Consistent with previous analysis, the black line in Fig.(\ref{sral}) stands for the Kerr case, but the black line in Fig.(\ref{srql}) stands for the R–N case with magnetic charge, and the values corresponding to the zero of the horizontal axis in these two black lines stand for the Schwarzschild case. These three data are consistent with previous calculations.

 From Figs.(\ref{sral})-(\ref{srql}), we can also find that, in comparison to the R-N and Kerr cases, the regular and rotating magnetic BH has smaller $\theta_\infty$ and $r_m$, but larger $s$. It is illustrated in these three figures that the rotating parameter $a$ has a greater influence on the observables than $q$ and $\lambda$, especially the change of $\theta_\infty$ with $a$, which can reach $50~\mu arcsecs$. At this stage, the observation accuracy of EHT has reached about $20~\mu arcsecs$, so this data can definitely be observed. However, the difference between the regular and rotating magnetic BH and Kerr BH is mainly determined by $q$ and $\lambda$, and $q$ has a maximum effect of $10~\mu arcsecs$ on $\theta_\infty$. Therefore, it may be a difficult task for future astronomical instruments to differentiate this BH from a GR one.

\newpage

\section{Summary}

In this paper, we study the strong gravitational lensing of the regular and rotating magnetic BH in non-minimally coupled EYM theory which characterized by rotating parameter $a$, magnetic charge $q$ of the Wu-Yang gauge field and EYM parameter $\lambda$. The black hole will reduce to Kerr case if we taking $\lambda=0$ and $q=0$, to R–N case with magnetic charge by setting $\lambda=0$ and $a=0$, to Schwarzschild case by letting $\lambda=0$, $q=0$ and $a=0$. By investigating the features of light propagation on the equatorial plane, we found that event horizon radius $r_h$,  photon sphere radius $r_{ps}$  and strong field limit coefficient $\bar{b}$ decrease, while $\bar{a}$ and deflection angle $\alpha(\theta)$ increase with the increase of any characteristic parameters of BH. By assuming that the massive compact object at the center of our galaxy can be described by this regular and rotating magnetic BH, we got the variation of observables in strong gravitational lensing, i.e., angular image position $\theta_{\infty}$ and relative magnification $r_m$ decrease while angular image separation $s$ increases with the increase of characteristic parameters.
We also showed that the rotating parameter $a$ has a greater influence on the strong gravitational lensing than magnetic charge $q$ and EYM parameter $\lambda$. The influence of $a$ is mainly reflected in the photon sphere radius $r_{ps}$ and deflection angle $\alpha(\theta)$, $r_{ps}$ for anterograde light rays ($a>0$) is always smaller than the retrograde ones ($a<0$),  but the change in deflection angle $\alpha (\theta )$ is opposite to $r_{ps}$.
That is to say the nearer the light is to the BH, the greater the deflection angle becomes.
These results will degenerate to that of the  Kerr case, R–N case with magnetic charge and Schwarzschild case when we take some specific values for the black hole parameters.
In addition, we used observables to quantify the influence of $q$ and $\lambda$, the biggest change in $\theta_\infty$ after being affected by $q$ is $10~\mu arcsecs$. Therefore, it may be a difficult task for current astronomical instruments to differentiate this BH from a GR one.

 \begin{acknowledgments}
{
{Thanks for the support of the  National Natural Science Foundation
of China under Grant No.11865018; the Henan Provincial Natural Science Foundation of China  under Grant No. 232300421351; the Talent Introduction Fund (Grant no. 2018BS042, 2020BS035) at Henan University of Technology.}}

\end{acknowledgments}

\textbf{Data availability Statement:} No data were generated or analyzed in the presented research.


\begin{thebibliography}{00}

 \bibitem{Abbott2016}
B. P. Abbott et al. (LIGO Scientific and Virgo Collaborations), Phys. Rev. Lett. \textbf{116}, 061102
(2016).

\bibitem{Abbott2016(2)}
 B. P. Abbott et al. (LIGO Scientific and Virgo Collaborations), Phys. Rev. Lett. \textbf{116}, 241103
(2016).

\bibitem{Abbott2017}
B. P. Abbott et al. (LIGO Scientific and Virgo Collaborations), Phys. Rev. Lett. \textbf{118}, 221101
(2017).

\bibitem{Abbott2017(2)}
 B. P. Abbott et al. (LIGO Scientific and Virgo Collaborations), Phys. Rev. Lett. \textbf{119}, 141101
(2017).


\bibitem{EHT} The Event Horizon Telescope Collaboration, K. Akiyama, A. Alberdi, et al.,
Astrophys. J. Lett, {\bf 875}, L1 (2019), arXiv:1906.11238.
\bibitem{EHT1} The Event Horizon Telescope Collaboration, K. Akiyama, A. Alberdi, et al., Astrophys. J. Lett. {\bf 875}, L2 (2019), arXiv:1906.11239.
\bibitem{EHT2} The Event Horizon Telescope Collaboration, K. Akiyama, A. Alberdi, et al., Astrophys. J. Lett. {\bf 875}, L3 (2019), arXiv:1906.11240.
\bibitem{EHT3} The Event Horizon Telescope Collaboration, K. Akiyama, A. Alberdi, et al., Astrophys. J. Lett. {\bf 875}, L4 (2019), arXiv:1906.11241.
\bibitem{EHT4}  The Event Horizon Telescope Collaboration, K. Akiyama, A. Alberdi, et al., Astrophys. J. Lett. {\bf 875}, L5 (2019), arXiv:1906.11242.
\bibitem{EHT5}  The Event Horizon Telescope Collaboration, K. Akiyama, A. Alberdi, et al., Astrophys. J. Lett. {\bf 875}, L6 (2019), arXiv:1906.11243.
\bibitem{EHT6} The  Event Horizon Telescope Collaboration, K. Akiyama, A. Alberdi, et al.,
 Astrophys. J. Lett. {\bf 930}, L17 (2022).

 \bibitem{Jing}
J. Jing, S. Long, W. Deng, M. Wang, and J. Wang,  
Sci. China, Phys. Mech. Astron. {\bf 65}, 100411  (2022), ArXiv: 2208.02420.

\bibitem{Jing1}
J. Jing, S. Chen, M. Sun, X. He,  M. Wang, and J. Wang,
Sci. China, Phys. Mech. Astron. {\bf 65}, 260411 (2022), ArXiv: 2112.09838.

\bibitem{Jing2023}J. Jing, W. Deng, S. Long and J. Wang,
Sci. China,  Phys. Mech. Astron. {\bf 66},   (2023).

\bibitem{Jing2021} Y. H. Zou, M. J. Wang and J. L Jing,
Sci. China, Phys. Mech. Astron. {\bf 64}, 250411 (2021).

\bibitem{Fang} W. Liu, X. Fang, J. Jing, A. Wang, 
Sci. China, Phys. Mech. Astron. {\bf 66}, 210411 (2023).

\bibitem{Chen} X. Zhou, S. Chen, J. Jing,
Sci. China, Phys. Mech. Astron. {\bf 65}, 250411 (2022).

\bibitem{Pan}
L. OuYang, D. Wang, X. Qiao, M. Wang, Q. Pan, J. Jing,
Sci. China, Phys. Mech. Astron. {\bf 64}, 240411 (2021).



\bibitem{Einstein} A. Einstein, Science {\bf 84} 506 (1936).

\bibitem{Walash} D. Walash, R. F. Carswell and R. J. Weymann,
 Nature {\bf279}, 381 (1979).

 \bibitem{kids} S. Li et al.,
 Astron. Astrophys., {\bf670}, A100 (2023).

 \bibitem{gaia} M. Jabłońska et al.,
 Astron. Astrophys., {\bf666}, L16 (2022).

  \bibitem{strides} T. Schmidt et al.,
 Mon. Not. R. Astron. Soc, {\bf518}, 1 (2023).

\bibitem{Hanson} D. Hanson et al.,
Phys. Rev. Lett, {\bf111}, 141301 (2013).

\bibitem{Courbin} Frédéric Courbin and Dante Minniti, Springer Berlin, Heidelberg
doi:10.1007/3-540-45857-3. ISBN 978-3-540-44355-1. ISSN 0075-8450.

\bibitem{Vir}  K. S. Virbhadra and G. F. R. Ellis,
 Phys. Rev. D {\bf 62}, 084003 (2000).

 \bibitem{Bozza3} V. Bozza, S. Capozziello, G. lovane, and G. Scarpetta,
Gen. Rel. and Grav. {\bf 33}, 1535 (2001).

\bibitem{Bozza} V. Bozza,
Phys. Rev. D {\bf 66}, 103001(2002).

\bibitem{Bozza2} V. Bozza,
Phys. Rev. D {\bf 67}, 103006(2003).

 \bibitem{schen3}
S. Chen, L. Zhang and J. Jing, Eur. Phys. J. C {\bf 78}, 981 (2018).

\bibitem{zhang}R. Zhang, J. Jing and S. Chen, Phys. Rev. D {\bf 95}, 064054 (2017).
\bibitem{zhang1}R. Zhang and J. Jing, Eur. Phys. J. C {\bf 78}, 796 (2018).

\bibitem{Azam} M. Azam, G. Abbas and S. Sumera,
Can. J. Phys. {\bf 95} 11, 1062-1067 (2017).

\bibitem{Geenner} H. F. M. Geenner, Living Rev. Relativity {\bf 17}, 5 (2014).

\bibitem{Hehl} F. W. Hehl and Yu. N. Obukhov, How does the electromagnetic field couple to gravity, in particular to metric, non-metricity, torsion, and curvature?, in Gyros, Clocks, Interferometers...: Testing relativistic gravity in space, edited by C. Lämmerzahl, C. W. F. Everitt, and F. W. Hehl, [Lect. Notes Phys. 562, 479 (2001)]

\bibitem{Hoissen} F. Müller-Hoissen,
Classical Quantum Gravity {\bf 5}, L35 (1988).

\bibitem{Balakin} A. B. Balakin, H. Dehnen and A. E. Zayats,
Phys. Rev. D  {\bf 76}, 124011 (2007).

\bibitem{Balakin2} A.B. Balakin and W.-T. Ni,
Classical Quantum Gravity {\bf 27}, 055003 (2010).

\bibitem{Balakin3} A. B. Balakin and A. E. Zayats,
Phys. Lett. B {\bf 644}, 294–298 (2007).

\bibitem{Balakin4} A. B. Balakin, J. P. S. Lemos and A. E. Zayats,
Phys. Rev. D {\bf 93}, 024008 (2016).

\bibitem{xie} F. Liu, Y. Mai, W. Wu and Y. Xie,
Phys. Lett. B {\bf 795}, 475–481 (2019).

\bibitem{Jawad} A. Jawad and M.U. Shahzad,
Eur. Phys. J. C {\bf 77}, 349 (2017).

\bibitem{Jawad2} A. Jawad and A. Khawer,
Eur. Phys. J. C {\bf 78}, 837 (2018).

\bibitem{Jusufi} K. Jusufi, M. Azreg-Aïnou, M.Jamil, S.Wei, W.Qiang and A.Wang,
Phys. Rev. D {\bf 103}, 024013 (2021).


\bibitem{Kala} S. Kala, H. Nandan and P. Sharma,
Eur. Phys. J. Plus 137-457 (2022).

\bibitem{Genzel} R. Genzel, F. Eisenhauer and S. Gillessen, Rev. Mod. Phys.  {\bf 82}, 3121 (2010).




\end{thebibliography}
\end{document}